\documentstyle[12pt]{article}
\font\titolo=cmbx12 scaled\magstep2
\font\titolino=cmbx12
\font\tsnorm=cmr12

\font\tscorsp=cmti10

\def\NPB{Nucl. Phys. }
\def\PLB{ Phys. Lett.  }  

\def\PRD{Phys. Rev.  }

\def\CMP{ Comm. Math. Phys. }
\def\AP{ Ann. Phys. }
\def\z{Z\kern -4.6pt Z}

\def\xp{x^+}
\def\l{\lambda}
\def\lq{\l^2}

\def\ha{{1\over 2}}

\def\d{\delta}
\def\i{\iota}

\def\o{\omega}
\def\p{\pi}

\def\s{\sigma}

\def\y{\eta}
\def\z{\zeta}

\def\P{\Pi}

\def\lie{{\cal L}}
\def\de{\partial}

\def\inf{\infty}

\def\mo{{-1}}
\def\ha{{1\over 2}}

\def\ds{ds^2=}

\def\g{\sqrt{-g}}

\def\d{\delta}
\def\e{\eta}
\def\eo{\e_0}
\def\m{\mu}
\def\n{\nu}

\def\ord#1{o\left(#1\right)}
\def\xo{\x^\perp}
\def\xp{\x^\parallel}
\def\ul{{1\over\l}}
\def\Ht{{\cal H}}
\def\Hx{{\cal H}_x}
\def\Pe{\P_\y}
\def\Ps{\P_\s}

\def\s{\sigma}
\def\x{\chi}
\def\ds{ds^2=}

\def\la{\l^2}
\def\i{\infty}
\def\be{\begin{equation}}
\def\ee{\end{equation}}
\def\bea{\begin{eqnarray}}
\def\eea{\end{eqnarray}}
\def\bc{\begin{displaymath}}
\def\ec{\end{displaymath}}
\def\lb{\label}
\def\ads{Anti-de Sitter }
\begin{document}
\pagestyle{empty}
\null
\vskip 5truemm
\begin{flushright}
INFNCA-TH9813\\
\end{flushright}
\vskip 15truemm
\begin{center}
{\titolo ENTROPY OF 2D BLACK HOLES}
\end{center}
\begin{center}
\titolo{FROM COUNTING MICROSTATES}
\end{center}
\vskip 15truemm
\begin{center}
{\tsnorm Mariano Cadoni$^{a,c,*}$ and Salvatore Mignemi$^{b,c,**}$}
\end{center}
%\smallskip
\begin{center}
{$^a$\tscorsp Dipartimento di Fisica,  
Universit\`a  di Cagliari,}
\end{center}
%\smallskip
\begin{center}
{\tscorsp Cittadella Universitaria, 09042, Monserrato, Italy.}
\end{center}
%\smallskip
%\smallskip
\begin{center}
{\tscorsp $^b$  Dipartimento di Matematica, Universit\'a  di Cagliari,}
\end{center}
%\smallskip
\begin{center}
{\tscorsp viale Merello 92, 09123, Cagliari, Italy.}
\end{center}
%\smallskip
\begin{center}
{\tscorsp $^c$  INFN, Sezione di Cagliari.}
\end{center}
\vskip 19truemm
%\baselineskip=2\normalbaselineskip
%%%%%%%%%%%%%%%%%%%%%%%%%%%%%%%%%%%%%%%%%%%%%%%%%%%%%%%%%%%%%%%%%%%%%%
%%		      abstract				       %%
%%%%%%%%%%%%%%%%%%%%%%%%%%%%%%%%%%%%%%%%%%%%%%%%%%%%%%%%%%%%%%%%%%%%%%
\begin{abstract}
\noindent
We present a microscopical derivation of the entropy of the black hole
solutions of the Jackiw-Teitelboim theory. We show that the asymptotic 
symmetry of two-dimensional (2D) Anti-de Sitter space is generated by 
a central
extension of the Virasoro algebra. Using a canonical realization of this
symmetry and Cardy's formula we calculate the statistical 
entropy of 2D black holes, which turns out to agree, up to a factor 
$\sqrt 2$,  
with the thermodynamical result.  
\end{abstract}
%%%%%%%%%%%%%%%%%%%%%%%%%%%%%%%%%%%%%%%%%%%%%%%%%%%%%%%%%%%%%%%%%%%%%%
%%			   End of abstract			       %%
%%%%%%%%%%%%%%%%%%%%%%%%%%%%%%%%%%%%%%%%%%%%%%%%%%%%%%%%%%%%%%%%%%%%%%
%%%%%%%%%%%%%%%%%%%%%%%%%%%%%%%%%%%%%%%%%%%%%%%%%%%%%%%%%%%%%%%%%%%%%%
%%			       Address				       %%
%%%%%%%%%%%%%%%%%%%%%%%%%%%%%%%%%%%%%%%%%%%%%%%%%%%%%%%%%%%%%%%%%%%%%%
\vfill
\begin{flushleft}
{\tsnorm PACS: 04.70. Dy, 04.50. +h  \hfill}
\end{flushleft}
%\begin{flushleft}
%{\tsnorm KEYWORDS: two-dimensional gravity models, black holes\hfill}
%\end{flushleft}
%\smallskip
%\vfill
\hrule
\begin{flushleft}
{$^*$E-Mail: CADONI@CA.INFN.IT\hfill}
\end{flushleft}
\begin{flushleft}
{$^{**}$E-Mail: MIGNEMI@CA.INFN.IT\hfill}
\end{flushleft}
\vfill
\eject
\pagenumbering{arabic}
\pagestyle{plain}

The idea of asymptotic symmetry  plays  an important role in the recent
developments in string theory and black hole physics. The Anti-de Sitter
(ADS)/conformal field theory (CFT) correspondence \cite{Wm}
is just one example of how asymptotic symmetries can be used to bring in touch
different theories in spacetimes of different dimensions. The conjectured
equivalence between Supergravity  on $D$-dimensional ADS space and  conformal 
field theory on the $D-1$-dimensional boundary is a very useful tool to gain
informations about  the nonperturbative regime of gauge  theories and 
to solve the 
problem of the microscopic interpretation of black hole entropy.

The previous ideas have found a  nice application for $D=3$. 
It is well known  since the work of Brown and Henneaux \cite{Bh} that the 
asymptotic symmetry group of ADS$_{3}$ is the conformal group in two 
dimensions. 
Using this result Strominger has calculated  the entropy  of the 
three-dimensional (3D) Ba\~nados-Teitelboim-Zanelli (BTZ)  black hole 
by counting
excitations of ADS$_3$ \cite {St}. A nice feature of this microscopical 
derivation of the 
black hole entropy is that it does not use string theory or supersymmetry,
but just general properties of 3D gravity. 
This fact makes the Strominger calculation of ref. \cite {St} more  similar 
to that of Carlip \cite{Car1}  
than to statistical derivations of black hole entropy that rely  both
on supersymmetry and string theory \cite{Sv}.

It looks very natural to try to apply the microstate counting procedure
of Strominger  to two-dimensional (2D) black hole solutions in ADS spacetime.
The simplest 2D gravity theory that admits ADS space as solution is the
Jackiw-Teitelboim (JT) model \cite{JT}. The 
JT model admits solutions that can be interpreted as 2D black holes
in ADS space and that behave very similarly to their four- and 
three-dimensional cousins. One can  associate with them a
Hawking temperature and a thermodynamical entropy \cite{CM}.
Moreover, at the semiclassical level  takes place the evaporation process,
 whose  Hawking radiation flux has been  already calculated \cite{CM}.

In this letter we present a microscopical derivation of the entropy of the
black hole solutions of the JT model. The approach used 
in ref. \cite{St} for the  3D case cannot be immediately extended to 
the 2D one.  
The obstruction is mainly due to the dimensionality of the $x\to \infty$
boundary of ADS$_2$,  which makes both the mathematical treatment and the 
physical interpretation of the results highly non trivial. 
For this reason  we  will present here only the main outcomes of our
investigation.
The details of the calculations and a thorough discussion of the physical
meaning of our results will be published elsewhere.

We compute the entropy of the JT black hole by counting
states on the one-dimensional, timelike, $x\to\infty$, boundary of ADS$_2$. 
To this end we first show how the $SL(2,R)$ isometry group  of ADS$_2$ can
be promoted to an asymptotic symmetry group on the boundary. This  asymptotic
symmetry group turns out to be generated by a central extension of the 
Virasoro algebra. Using a canonical realization of the asymptotic symmetry, 
we calculate the central charge $c$ of the algebra. Applying Cardy's 
formula \cite{Ca} for the asymptotic density of states, we calculate the 
statistical entropy of the JT black hole reproducing, up to a factor 
$\sqrt 2$ the thermodynamical result.

The JT model is described by the action
\be\lb{e1}
A={1\over2}\int\g\ d^2x\ \e\left(R+2\lq\right),    
\ee
where $\l$ is the 2D cosmological constant and $\e$ is a scalar
field related to the usual definition of the dilaton $\phi$ by 
$\e=\exp(-2\phi)$.
The theory admits solutions that can be interpreted as 
2D black holes in ADS space, which in a Schwarzschild gauge take the 
form \cite{CM}:     
\be\lb{e2}
\ds-(\l^2x^2-a^2)dt^2+(\l^2x^2-a^2)^{-1}dx^2,\qquad \e=\eo \l x,
\ee
where $\eo$ is an integration constant and $a^2$ is related to mass $M$ of
the black hole by
\be\lb{e3}
M={1\over 2} \eo a^2\l.
\ee
Two-dimensional dilaton gravity does not allow for a dimensionful analog 
of the Newton constant. However, it is evident from the action 
(\ref{e1}) that the inverse of the scalar field $\e$ represents the
(coordinate-dependent) coupling constant of the theory, 
whereas the inverse of
the integration 
constant
$\eo$  plays the role of a dimensionless 2D Newton constant.

All the solutions (\ref{e2}) are locally \ads, but have different global 
properties.
In particular, we consider the $a=0$ solutions 
(which  following the notation of
ref. \cite{CM} will be denoted by ADS$^0$) as the ground state of the model.
ADS$^0$ is not geodesically complete and differs globally from full 
2D ADS space (the $a^2=-1$ solution in eq. (\ref{e2}) ) \cite{CM}.
A similar phenomenon occurs also for the 3D BTZ black hole solutions.

Using standard arguments one can easily calculate the thermodynamical 
parameters
associated to the black hole  (\ref{e2}).
For the entropy $S$ we have \cite{CM}:
\be\lb{e4}
S=4\pi \sqrt {\eo M\over 2\l}=2\pi {\e_h},
\ee
where $\e_h$ is the value of the scalar field  at the horizon.
In two spacetime dimensions we do not have an area law for the black hole 
entropy.
However the second equality in eq. (\ref{e4}) can be interpreted as
a generalization to 2D of the Bekenstein-Hawking entropy.
This follows simply from the fact that  according to eq. (\ref{e2}),
$\e$ is nothing but the "radial" coordinate of the 2D space.

The \ads space is invariant under the $SO(1,2)\sim SL(2,R)$
group of isometries which, in the case of ADS$^0$, are generated by the three
Killing vectors
\be\lb{e5}
^{(1)}\x={1\over \l}{\de\over\de t},\qquad\qquad^{(2)}\x=t{\de\over\de t}-
x{\de\over\de x},\qquad\qquad^{(3)}\x=\l \left(t^2+{1\over\l^4x^2}
\right){\de\over\de t}-2\l tx{\de\over\de x}.
\ee

The asymptotic symmetries are best investigated in the hamiltonian formalism.
With the parametrization
\be\lb{e6}
\ds-N^2dt^2+\s^2(dx+N^xdt)^2,
\ee
the hamiltonian of the JT theory reads \cite{Kuchar}
\be\lb{e7}
H=\int dx(N\Ht+N^x\Hx).
\ee
$N$ and $N^x$ act as usual as Lagrange multipliers enforcing the constraints,
\bea\lb{e8}
\Ht&=&-\Pe\Ps+\s^\mo\y''-\s^{-2}\s'\y'-\l^2\s\y=0,\nonumber\\
\Hx&=&\Pe\y'-\s\Ps'=0,
\eea
where
\be\lb{e9}
\Pe=N^\mo(-\dot\s+(N^x\s)'),\qquad\qquad \Ps=N^\mo(-\dot\y+N^x\y'),
\ee
are the momenta conjugate to $\y$ and $\s$, respectively.
A dot denotes derivative with respect to $t$ and a prime with respect to $x$.

In case of non-compact spacelike surfaces, however, it is well known that,
in order to have well defined variational derivatives, one must add
to the hamiltonian a surface term $\d J$, which in general depends on the
boundary conditions imposed on the fields \cite{Regge}.
In our case, the boundary reduces to a
point and the variation  $\d J$ must be given by
\be\lb{f1}
\d J=-\lim_{x\to\inf}[N(\s^\mo\d\y'-\s^{-2}\y'\d\s)-N'(\s^\mo\d\y)+
N^x(\Pe\d\y-\s\d\Ps)].
\ee
Using suitable boundary conditions, this can be written as a total variation at
infinity of a functional $J$.

We have now to fix the boundary conditions at spatial infinity such that  
the metric behaves asymptotically as that of ADS$_{0}$  and to study 
under which
transformations they are preserved. We require that, for $x\to\inf$
\be\lb{f2}
g_{tt}\sim -\l^2x^2+o(1),\qquad g_{tx}\sim\ord{1\over x^{3}},\qquad
g_{xx}\sim{1\over\l^2x^2}+\ord{1\over x^4}.
\ee
Actually, in order to enforce \ads behaviour at infinity, 
one could choose milder
asymptotic conditions. However, our stronger conditions are needed in
order to have well-defined charges $J$.
The asymptotic conditions (\ref{f2}) imply
\be\lb{f3}
\s\sim{1\over\l x}+\ord{1\over x^3},\qquad N\sim \l x+\ord{1\over x},
\qquad N^x\sim \ord{1\over x}.
\ee
Imposing that the asymptotic form (\ref{f2}) of the metric is conserved 
under the
action of the Killing
vectors $\x^\m$, one obtains that these must have the form:
\be\lb{f4}
\x^t=T(t)+{1\over 2 \l^{4}}{d^{2} T(t)\over dt^2} {1\over x^{2}}+
\ord{1\over x^4},
\qquad\qquad\x^x=- {dT(t)\over dt}\, x+\ord{1\over x},
\ee
where $T$ is an arbitrary function of $t$.
Diffeomorphisms with $T=0$ fall off rapidly as $x\to \i$. They represent "pure"
gauge transformations.

One still has to consider how the transformations (\ref{f4}) affect the dilaton.
The variation of a scalar field $\y$ is given by $\lie_\x\y=\x^\m\de_\m\y$,
which is asymptotically $o(x)$ for $\y$ of the form (2), and hence of the same
order as the field itself. This is quite disturbing, but is an inescapable
consequence of the scalar nature of the dilaton, and is also in accordance with
the fact that $\y$ is defined up to  the scale factor 
$\eo$ by the field equations.
The previous considerations together with eq. (\ref{e9}) permit to fix the
asymptotic behaviour of the remaining canonical variables:
\be\lb{f6}
\y\sim o(x),\qquad\qquad\Ps\sim o(1),\qquad\qquad\Pe\sim o(x^{-4}).
\ee

We can now write down the algebra generated by the asymptotic symmetries
(\ref{f4}).
Since the \ads space has a natural periodicity in $t$, it is convenient to
expand the function $T(t)$ in a Fourier series in the interval 
$0<t<2\p/\l$.
The generators of the asymptotic symmetries read then,
\bea\lb{f7}
A_k&=&\ul\left(1-{k^2\over2\l^2x^2}\right)\cos(k\l t){\de\over\de t}+
kx\sin(k\l t){\de\over\de x},\nonumber\\
B_k&=&\ul\left(1-{k^2\over2\l^2x^2}\right)\sin(k\l t){\de\over\de t}-
kx\cos(k\l t){\de\over\de x},
\eea
where $k$ is an integer. The generators satisfy the commutation relations,
\bea\lb{f8}
& &[A_k,A_l]=\ha(k-l)B_{k+l}+\ha(k+l)B_{k-l},\nonumber\\
& &[B_k,B_l]=-\ha(k-l)B_{k+l}+\ha(k+l)B_{k-l},\nonumber\\
& &[A_k,B_l]=-\ha(k-l)A_{k+l}+\ha(k+l)A_{k-l},
\eea

In the hamiltonian formalism, the symmetries associated with the Killing
vectors $\x^\m$ are generated by the
phase space functionals $H[\x]$, defined as
\be\lb{h1}
H[\x]=\int dx \big(\xo\Ht+\xp\Hx\big)+J[\x],
\ee
where $\xo=N\x^t$, $\xp=\x^x+N^x\x^t$, and the surface term $J[\x]$ can be
interpreted as the charge associated with the symmetry generator $\x^\m$.
In view of the boundary conditions discussed above and 
adjusting the 
arbitrary constant  so that $J$ vanishes for ADS$^0$,
the functional $J[\x]$ can be written in finite form as
\be\lb{h2}
J[\x]=\lim_{x\to\inf} \eo \left [- (\l x)\xo (\y'-\l)+(\l x)
{\partial \xo\over \partial r}(\y-\l x)+ {\l^{4}x^{3}\over 2}  \xo 
\left ( g_{xx}- {1\over \la 
x^{2}}\right) +{1\over \l x} \xp \Ps \right].
\ee

In general, the Poisson bracket algebra of $H[\x]$ yields a projective 
representation of the asymptotic symmetry group \cite{Bh}:
\be\lb{h3}
\{H[\x],H[\o]\}=H[[\x,\o]] +c(\x,\o),
\ee
where $c$ is the central charge of the algebra. 
By enforcing the constraints $\Ht_\n=0$ the charges $J[\x]$ give themselves
a realization of the asymptotic symmetry group through  the
Dirac bracket, so that
\be\lb{h4}
\{J[\x],J[\o]\}_{DB}=J[[\x,\o]] +c(\x,\o).
\ee
In the case of three-dimensional \ads space, the previous arguments give  
a simple way to calculate the central charge of the algebra \cite{Bh}.
One just needs to observe that the surface deformation algebra 
$[\x,\o]_{SD}$  is isomorphic to the  algebra of the asymptotic symmetries
and that the
variation of $J[\x]$ under surface deformations is given by the Dirac bracket,
\be\lb{h5}
\delta_\o J[\x]=J[[\x,\o]] +c(\x,\o).
\ee
By evaluating the previous equation for $ADS^{0}$, one finds that the central
charge $c(\x,\o)$ is just given by the charge $J[\x]$ evaluated on the 
surface deformed by $\o$.

In the case of 2D \ads space, however, the previous calculation method cannot
work, at least in the form described above. In fact, being the boundary
a point,  the functional derivatives appearing in the Poisson bracket
(\ref {h3}) can be defined only for pure gauge transformations, for which 
the charge $J[\x]$ vanishes. Moreover, the Dirac brackets (\ref{h4}) have no
meaning as
long as the $x\to \i$ boundary is  a point. As a consequence, the 
surface deformation algebra has no definite action on the charges $J[\x]$,
and eq. (\ref{h5}) cannot be used to calculate the central charge.

The simplest way to cure the disease is to define the time-independent 
charges
\be\lb{h6}
\hat J[\x]={\l\over 2\pi} \int_0^{2\pi/\l} dt\, J[\x].
\ee
The functional derivatives  of $\hat J[\x]$ can be easily  defined,
so that the Dirac bracket algebra $\{\hat J[\x],\hat J[\o]\}_{DB}$ 
has now a meaning. One can also  verify that the action 
of the surface deformation on the charges $\hat J[\x]$ gives a realization
of the  algebra (\ref{f8}).
Let us comment briefly on the physical meaning of the charges $\hat J$.
Apart from $J[A_0]$, which gives the mass $M$ of the solution, the other
charges $J[A_k]$ are in general time-dependent. This means that besides 
the mass there are  no conserved quantities. This fact is strongly related to
the presence of the dilaton and its behaviour under the 
transformations (\ref {f4}). On the other hand the charges 
$\hat J$ represent a sort of
averaged charges that can be used to give a canonical representation
of the  algebra (\ref {f8}).

We can now easily calculate the central charges $c$.
We just need to use in eq. (\ref {h5})  the charges $\hat J$ instead of 
$J$. One gets,
\be\lb{h7}
c(A_k,A_l)=c(B_k,B_l)=0, \qquad c(A_k, B_l)= \eo k^3 \d_{|k|\,|l|}.
\ee
Defining new generators $L_k=-(B_k-iA_k)$, and shifting $L_0$ by  a 
constant, one obtains the Virasoro algebra,
\be\lb{f9}
[L_k,L_l]=(k-l)L_{k+l} + {c\over 12}(k^3-k)\d_{k+l},\quad c=24\e_0.
\ee  

To calculate the entropy of a generic black hole solution of mass $M$ in 
terms of states living on the boundary, we just need to
use Cardy's formula for the asymptotic density of states:
\be\lb{g1}
S=2\pi \sqrt{c\,l_0\over 6},
\ee
where $l_0$ is the eigenvalue of the Virasoro generator $L_0$, which
for a black hole of mass $M$ is given by
\be\lb{g2}
l_0= {M\over\l}.
\ee
Inserting eq. (\ref{g2}) and the value of the central charge $c$ given by eq.
(\ref{f9}) into eq. (\ref {g1}), we find for the statistical entropy, 
\be\lb{g3}
S= 4\pi \sqrt {\eo M\over \l},
\ee
which agrees, up to a factor $\sqrt 2$,  with the thermodynamical 
result (\ref{e4}).
The lack of knowledge about the theory on the boundary renders  
difficult  explaining  this discrepancy 
between the statistical  and the thermodynamical result. Nevertheless,  
a simple 
explanation of the factor  $\sqrt{2}$ can be found  if one considers 
the model (\ref{e1}) as a circular symmetric dimensional reduction of 
three-dimensional 
gravity, with  the field $\e$ parametrizing the radius of the circle.
Using the notation of ref. \cite{St}, the 2D dilaton gravity action 
can be obtained from the 3D one by the ansatz
\be\lb{g4}
ds^{2}_{(3)}= ds^{2}_{(2)}+ 16G\e^{2}d\varphi^{2},
\ee
where $G$ is the 3D Newton constant and $0\le\varphi\le 
2\pi$.
In this context the 2D black hole (\ref{e2}) can be 
considered as the dimensional reduction of the $J=0$ (zero angular momentum) 
BTZ black hole. Simple calculations show that both the mass and the 
thermodynamical entropy of the BTZ black hole agree with our 2D results.
The same is not true for the statistical entropy. From 
the 3D point of view we have contributions to the mass of the black 
hole coming from both the right- and left-movers oscillators of 
the 2D conformal field theory living on the boundary of ADS$_{3}$. 
Because $J=0$ implies that the number of right-movers equals that of 
left-movers, we have $l_{0}=M/2\l$, which inserted in the Cardy's 
formula reproduces the thermodynamical entropy (\ref{e4}).
From the 2D point of view only oscillators of one sector contribute 
to the mass of the black hole giving $l_{0}=M/\l$ and the statistical 
entropy (\ref{g3}). These results are in accordance with those obtained 
by Strominger in a recent paper \cite {st1}, where ADS$_{2}$ is 
generated as the near-horizon, near-extremal limit  of ADS$_3$.
At first sight  this seems to imply that there is no intrinsically 2D 
explanation of the statistical entropy of 2D black holes.
This is certainly true as long as  the field
$\e$ is interpreted as the radius of the internal circle, because the 
$x\to\infty$ boundary of ADS$_2$ corresponds to the region $\e\to \infty$, 
where   
the space decompactifies and the 2D theory becomes intrinsically 3D. 

The previous considerations do not apply when ADS$_2$ arises as 
near-horizon geometry of higher dimensional black holes with no 
intermediate ADS$_3$  geometry  involved. We do not have a complete
explanation of the factor  $\sqrt 2$ in this case. In our opinion what is
needed in order to find an explanation of this discrepancy 
is a complete understanding of the role played in our derivation by the 
global topology of ADS$_2$. Full ADS$_2$ has a cylindrical topology 
with two  disconnected timelike boundaries. This fact plays a crucial 
role in ref. \cite{st1} because it  makes the string theory on ADS$_{2}$
a theory of open strings. By studying the black hole solutions of
the JT theory we are forced to cut the spacetime on the $x=0$ 
``singularity'', so that only one timelike boundary of  full ADS$_{2}$
is available. It seems to us  that a thorough understanding of the 
statistical entropy of 2D black holes will be at hand only when 
this point will be fully clarified.

Our derivation of the statistical entropy of 2D black holes, though very simple
and elegant, has the same drawbacks as the derivation of Strominger
\cite{St} (for a critical review see ref. \cite{Car2}).
In particular remains open the question about the origin and the location
of the relevant degrees of freedom on the boundary, whose number of 
excitations  account for  the entropy of the  black hole. 
In our case, the nature of these degrees of freedom is even more mysterious 
than in the 3D case. Even though one has no explicit description of the 
degrees of freedom that are responsible for the entropy of the BTZ black
hole, the underlying field theory is  well known, being
2D conformal field theory with given central charge.
For 2D black holes, instead,
we know very little about the theory that should describe
the excitations on the boundary. The one-dimensional nature of the latter 
implies that we are dealing with some kind of particle quantum mechanics,
rather than  quantum
field theory. The quantum mechanical system, whose states span a representation
of the Virasoro algebra (\ref{f9}) is most likely a very unconventional one.
In this context the implementation  of the ADS/CFT correspondence in the 
2D case could help to shed light on the nature of this quantum mechanical
system.  On the other hand 
the fact that one can use particle quantum mechanics 
(even though in a still mysterious form) to explain the entropy of 2D
black holes  seems to us a very exciting possibility.

\begin {flushleft}
{\titolino Note added}
\end {flushleft}
After this manuscript was completed we became aware of the existence of the
paper of ref \cite{Ho}, in which the asymptotic symmetries of 2D \ads space
are  discussed.

\end{document}